# Transmission Lines Positive Sequence Parameters Estimation and Instrument Transformers Calibration Based on PMU Measurement Error Model

**Chen Wang[1, 2], (Student Member, IEEE), Virgilio A. Centeno[1], (Senior Member, IEEE), Kevin D. Jones[2], (Member, IEEE), and Duotong Yang[2], (Member, IEEE)**

[1]The Bradley Department of Electrical and Computer Engineering, Virginia Polytechnic Institute and State University, Blacksburg, VA 24061 USA
[2]Dominion Energy Virginia, Richmond, VA 23220 USA

Corresponding author: Chen Wang (e-mail: chenwang@vt.edu).

This work was supported by the Virginia Tech Open Access Subvention Fund

**ABSTRACT** Phasor Measurement Unit measurement data have been widely used in nowadays power system applications both in steady state and dynamic analysis. The performance of these applications running in utilities' energy management system depends heavily on an accurate positive sequence power system model. However, it is impractical to find this accurate model with transmission line parameters calculated directly with the PMU measurements due to ratio errors brought by instrument transformers and communication errors brought by PMUs. Therefore, a methodology is proposed in this paper to estimate the actual transmission lines parameters throughout the whole system and, at the same time, calibrate the corresponding instrument transformers. A PMU positive sequence measurement error model is proposed targeting at the aforementioned errors, which is applicable to both transposed and un-transposed transmission lines. A single line parameters estimation method is designed based on Least Squares Estimation and this error model. This method requires only one set of reference measurements and the accuracy can be propagated throughout the whole network along with the topology acquired by the introduced Edge-based Breadth-first Search algorithm. The IEEE 118-bus system and the Texas 2000-bus system are used to demonstrate the effectiveness and efficiency of the proposed method. The potential for deployment in reality is also discussed.

## NOMENCLATURE

| | |
|---|---|
| PMU | Synchronous phasor measurement unit |
| EMS | Energy management system |
| TL | Transmission line |
| CT | Current transformers |
| PT | Potential transformers |
| PS | Positive sequence |
| LSE | Least squares estimation |
| ADC | Analog to digital convertor |
| $R$ | Transmission line resistance |
| $X$ | Transmission line reactance |
| $y$ | Transmission line shunt susceptance |
| $Z$ | Transmission line impedance |
| $V_i$ | True PS voltage phasor on bus $i$ |
| $I_i$ | True PS current phasor on bus $i$ |
| $V_{iM}$ | Measured PS voltage phasor on bus $i$ |
| $I_{iM}$ | Measured PS current phasor on bus $i$ |
| $REV_i$ | Ratio error on $V_i$ brought by PT |
| $REI_i$ | Ratio error on $I_i$ brought by CT |
| $REV_{ph}$ | Single phase ratio error on voltage measurement, $ph \in \{A, B, C\}$ |
| $r_{M, ph}$ | Magnitude of single-phase ratio error |
| $r_{A, ph}$ | Phase angle of single-phase ratio error |
| $V_{PM}$ | Measured PS voltage |
| $V_{ph}$ | True single phase voltage phasor |
| $V_P$ | True PS voltage phasor |

| $REV_P$ | PS ratio error on voltage measurement |
| $REV_{P,ut}$ | PS ratio error on voltage measurement of un-transposed TL |
| $\beta_{ph}$ | Complex value indicating single-phase deviation from PS |
| $\mathbf{Z}$ | Impedance matrix of one single TL |
| $KV_i$ | Correction factor of $REV_i$ |
| $KI_i$ | Correction factor of $REI_i$ |
| $\varepsilon_{V_{iM}}$ | PS quantization error on $V_{iM}$ |
| $V_{iM.q}$ | Actual quantized PS voltage measurement on bus $i$ output by PMU |
| $I_{iM.q}$ | Actual quantized PS current measurement on bus $i$ output by PMU |
| $K_{real}$ | Real part of correction factors |
| $K_{imag}$ | Imaginary part of correction factors |
| $\mathbf{l}_{real}^{(true)}$ | Vector of real parts of accurate current measurements |
| $\mathbf{l}_{imag}^{(true)}$ | Vector of imaginary parts of accurate current measurements |
| $l_{real}$ | Vector of lower limits on $K_{real}$ |
| $u_{real}$ | Vector of upper limits on $K_{real}$ |
| $l_{imag}$ | Vector of lower limits on $K_{imag}$ |
| $u_{imag}$ | Vector of upper limits on $K_{imag}$ |
| $\mathbf{Z}_{abc}$ | Impedance matrix of example un-transposed TL (three-phase unbalanced) |
| $\mathbf{y}_{abc}$ | Susceptance matrix of example un-transposed TL (three-phase unbalanced) |

## I. INTRODUCTION

Synchronous Phasor Measurement Units have been widely deployed and utilized in today's high voltage transmission power grids. With high reporting rates and accurate GPS timestamps, these devices enable real-time system monitoring and analysis[1]. Applications are widely deployed based on PMU measurements for system modeling[2], fault locating[3], and state estimation[4], etc. The fundamental data provided by PMUs are voltage and current measurements obtained mostly from protection-type transducers: CTs and PTs installed throughout the network. Their measuring process of scaling down high voltage/current will inevitably introduce ratio errors. These errors, described in the IEEE Std C57.13 standards [5], distort the information of the system's instantaneous operating conditions illustrated by the measurements. One of the practical ways to resolve the ratio error is to utilize revenue type CTs/PTs. However, they are too costly to be applied across the entire system.

Furthermore, most utilities only collect PS data from installed PMUs. These PS measurements are computed from three-phase CT/PT measurements affected individually by the aforementioned ratio errors. In addition, due to the existence of un-transposed TLs, three-phase unbalance also has a negative impact on the accuracy of the measurements. This paper intends to resolve this error issue by developing a software-based method to find out the errors within the PS PMU measurements and to further estimate the system TL parameters for a more accurate system model.

There exist several studies focusing on the errors in PMU measurements. Reference [6] provides a comprehensive study to estimate PMU measurement errors using a nonlinear optimal estimation theory. The impact of the PMU measurement errors on the system dynamic load model is described and the stochastic distribution, as well as the frequency spectrum of the error, are analyzed in [7]. In reference [8], the PMU measurements errors are found to influence the system modeling in the steady state. To resolve this problem, the LSE is used to conduct calibration considering different PMU installation cases. Researchers also proposed estimating the CT and PT's ratio error using PMU measurements [9]. Pal et al., used a three-phase line model to calibrate the potential transformers with more accurate measurements [10]. However, most of the studies take PMUs as perfect information transferring media and ignore the loss of accuracy due to analog-to-digital conversion.

The researchers mentioned above also assumed the TL parameters are accurate and invariant from the time of installation. However, the line parameters are significantly influenced by operation conditions and environment. Zhang et al.[11] recognized this effectiveness, proposed a method to identify the suspicious erroneous line parameters and then estimate these parameters as states. However, this method highly depends on the accuracy of the errors identification. There have also been other methods proposed to estimate transmission line parameters considering PMU errors from perspectives of both transmission level [12] and distribution level [13]. Reference [14] introduces a method to estimate one single TL's parameters using recursive regression based on a Kalman filter. Reference [15] introduced a novel estimator to estimate the three-phase parameters of a single TL. An off-line methodology has been proposed in [16], which can estimate the single TL parameters with only one side of the line equipped with one PMU. Wu, Zora, and Phadke [17] have proposed an algorithm attempting to estimate both the three-phase line parameters and the ratio errors of the CTs/PTs. However, this method is aimed at perfectly balanced three-phase system and does not provide an effective method for the topological measurement propagation. More importantly, the paper does not consider the PMU quantization error and simply uses CTs/PTs ratio errors as the error of the PMU measurements. Researchers in [18] focused on un-transposed TL's parameters estimation. They combined the system state estimation and the three-phase line parameters real-time tracking. Gaussian white noise is used to simulate the PMU measurement errors. The CTs/PTs calibration and TL parameters estimation problem is also considered in [19]. Researchers derived the estimation problem by taking CT/PT errors as probabilistic variables. While in the process of extending the estimation to the whole network, the researchers used regular LSE method to find the current measurements errors and did not consider the irregular scenarios like parallel transmission lines.

The aim of this paper is to provide power utilities equipped with a limited number of revenue CTs and PTs with a method to find more accurate PS parameters of the TLs throughout the system and to calibrate all the non-revenue CTs and PTs with data analytic automation process. The whole process starts with revenue transducers installed on one single bus in the system and the estimation of line parameters and current/voltage measurement errors is conducted on TLs one by one throughout the grid.

The main contributions of this paper include:

1) A more realistic PMU PS measurement error model was formulated and deployed other than the normal distribution assumption on errors used in previous state/parameters estimation researches[20], [21]. The PMU measurement errors for PS voltages and currents measurements are analyzed based on which this model is proposed for further more realistic estimation method construction.

2) A novel algorithm is proposed to estimate PS parameters of transposed/un-transposed TLs and ratio errors of the corresponding CTs and PTs. Unlike [12], [22], the PS line parameters are estimated without directly using the LSE method but integrating a novel technic to consider more detailed PMU measurement errors characteristics.

3) A set of more adaptive propagation methods than [17], [19] are proposed to transmit the accuracy to the consecutive lines as the reference of the estimation of their parameters. The proposed method considers more comprehensive scenarios like parallel lines to make the propagation method more robust.

4) The edge-based breadth-first search algorithm is proposed to identify the sequence of the TLs to be estimated. As in previous studies [15], [17], the accuracy propagation methods were usually not discussed in detail. The introduction of this method provides a systematic way to find the system topology to propagate the accuracy acquired with the single TL estimation throughout the whole network.

This paper is organized as follows: Section II presents the TL model and PMU measurement error model utilized in the following analysis; Section III describes the methodology of estimating PS line parameters of single TLs, transducers calibration, and the propagation algorithms; Section IV presents the case study results; Section V concludes the article.

## II. TRANSMISSION LINE AND PMU MEASUREMENT ERROR MODEL

The actual estimation is conducted on every single TL, which is modeled with the widely used PS π-section model. The CTs/PTs, with PMUs, are placed on both ends of the line. For better estimation effectiveness, the characteristics of the error integrated by these devices are to be analyzed. The knowledge of these characteristics forms the foundation for further methodology establishment.

### A. Transmission Line π-section Model

The widely used TL model consists of PS line parameters: $R$, $X$, and $y$, as is shown in Figure 1. In the figure, it can be seen that the measured values of voltages and currents from both ends of the TL are different from the actual values. The TL impedance is $Z = R + jX$.

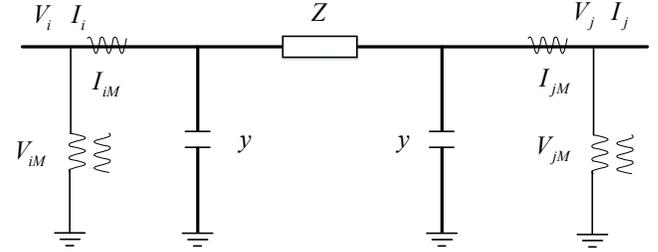

**Figure 1.** Transmission line π-section model and CT/PT measurements.

### B. PMU Measurements Error Model

The PMU measurements utilities use are mostly in PS. These data are transformed from three-phase measurements and quantized by ADC inside the PMUs. This conversion process can cause loss of information. On the other hand, the accuracy of these measurements also depends on the original measuring devices (CTs/PTs). Therefore, it is important to find out the characteristics of these errors along with the data flowing process. Given knowledge of these error features, the basis is built for more realistic PMU measurement simulation and further parameter estimation.

#### 1) MEASUREMENTS DATA FLOW

As shown in Figure 2, the actual voltages and currents on every single TL are measured by the CTs/PTs. Because of the core excitation current in the scaling down process, ratio errors are introduced. After receiving the signals from CTs and PTs, PMUs quantize each phase data and compute the PS voltage or current phasors, which also causes loss of precision.

Both the three-phase and PS ratio errors are complex values. The relationship among true values, measurements, and ratio errors on bus $k$ is shown in (1).

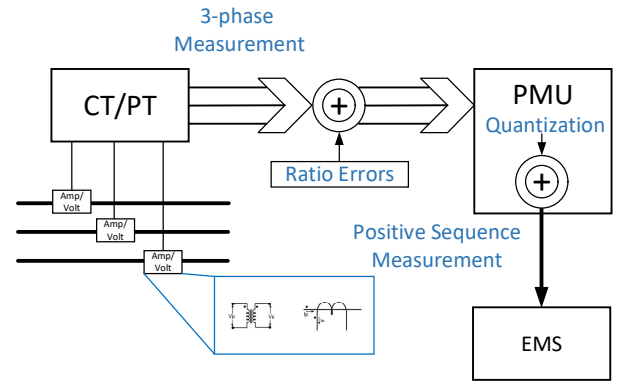

**Figure 2.** Voltage and current measurement data flow.

$$\begin{cases} V_{kM} = REV_k \times V_k \\ I_{kM} = REI_k \times I_k \end{cases} \quad (1)$$

For each phase, there is a different ratio error. Since these errors mainly depend on the CTs/PTs hardware, it can be assumed that during the time period of concern, such ratio errors remain fixed. According to[18], the ratio errors for each single phase is as follows:

$$REV_{ph} = r_{M,ph} \times e^{jr_{A,ph} \times \pi/180} , \quad ph \in \{ A, B, C \} \quad (2)$$

The PMUs take the individual three-phase values as input and computes PS values. Therefore, without losing generality, taking voltage as an example, the PS voltage measurements would be as follows:

$$V_{PM} = \frac{1}{3} \left( REV_A \cdot V_A + \alpha \cdot REV_B \cdot V_B + \alpha^2 \cdot REV_C \cdot V_C \right) \quad (3)$$

where, $\alpha = e^{j2\pi/3}$ .

If the three-phase voltage measurements are balanced, the PS ratio error, $REV_P$, can be computed as

$$REV_P = \frac{V_{PM}}{V_P} = \left( \frac{1}{3} \right) \left( REV_A + REV_B + REV_C \right) \quad (4)$$

However, in reality, there do exist some cases where TLs are un-transposed, which can cause three-phase imbalances in the measurements. Under such circumstances, the three-phase deviations from the PS are introduced and thus the PS ratio error becomes:

$$REV_{P,ut} = \frac{\beta_A \cdot REV_A + \beta_B \cdot REV_B + \beta_C \cdot REV_C}{\beta_A + \beta_B + \beta_C} \quad (5)$$

It is to be noted that as the three phases' ratio errors and the transposing condition of each TL remain constant, there is a constant PS ratio error for each set of three-phase PMU measurements.

### 2) PMU QUANTIZATION ERROR

The quantization behavior of the three-phase PMU output data is described in [10][23]. All three-phase voltage and current data received by PMUs are quantized by certain scales due to the analog-to-digital conversion [24]. Since the PS measurements are directly transformed from the quantized three-phase data, the quantization errors are also indiscriminately integrated. Given that many estimators deployed in utilities only consider PS network, these errors can lead to inaccurate or even incorrect system states and parameters identification.

The PS measurement errors caused by the three-phase quantization cannot be simply taken as an additive error because of the PS transformation process. Here, the simulated three-phase voltage data are quantized and taken as an example to demonstrate the effect of the PS transformation after quantization. 7,500 times of power flow are conducted based on an equal number of system load statuses in one morning load pick-up hour chronologically. The three-phase voltage phasors on one single bus are collected. The quantization errors of each single-phase data are simply the differences between the original phasors and the quantized ones. Real and imaginary parts are quantized separately. The voltage level is 345KV and the quantization scale which is 12.16V. For the PS, the true PS phasors are transformed with true three-phase data and the quantized PS phasors are transformed with quantized three-phase data. The PS quantization errors are the differences between them.

The histogram of the quantization errors of three-phase and PS voltage data are shown in Figure 3. The histogram bin is set to be 1/20 of the quantization scale. As can be seen in Figure 3, the quantization process causes losing accuracy of the three-phase measurements due to the quantization errors. For both real and imaginary parts of the three-phase voltage data ("VA real quant error" and so on), the quantization errors are uniformly distributed in general. As for the PS data ("VP real quant error" and so on), the errors show bell shapes brought by the sum of three uniform distributions [25]. Notably, the expectation of this error is close to zero and therefore unbiased. The existence of quantization errors in the PS measurements clearly shows that these errors cannot be ignored within any estimation process using these data. Therefore, in the following section, a series of estimation methods is designed to resolve these errors taking advantage of their unbiased characteristic.

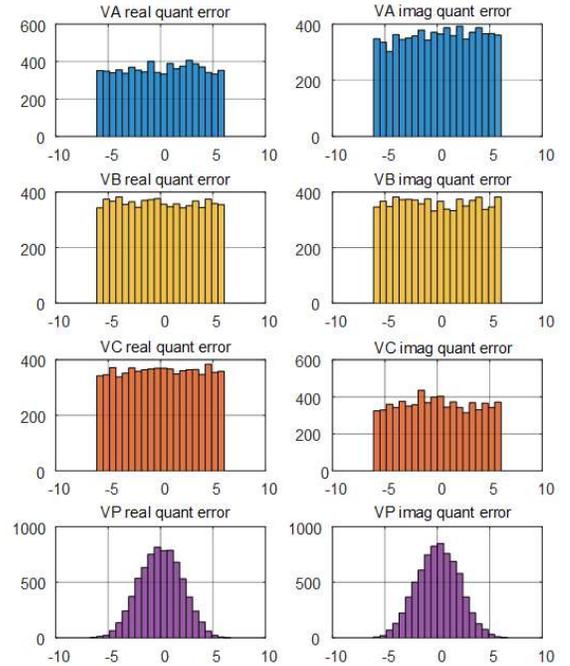

**Figure 3.** Simulated Positive sequence and three-phase voltage quantization errors distribution histogram.

## III. LINE PARAMETERS ESTIMATION METHODOLOGY

With the knowledge of the PMU measurement errors characteristics, a linear estimation methodology based on the LSE is proposed to estimate the line parameters and to calibrate the transducers. The methodology consists of two parts: single TL estimation and accuracy propagation. The process starts with one single TL in the network. On its "*From bus*", it is assumed that there is one set of revenue class CTs and PTs installed. The voltage and current measurements on this bus are accurate and are taken as the reference for the whole estimation process.

Given the accurate measurements on the "*From bus*", one can estimate line parameters and the ratio errors within the measurements on the "*To bus*" at the same time. Different from [17], during this single line estimation, the PMU quantization errors are considered.

Since the "*To bus*" of the currently concerned line is also the "*From bus*" of the consecutive line, detailed techniques are developed to propagate the accuracy to the next TL for parameter estimation. Thus, following the connected system topology, parameters of all the concerned lines can be estimated.

### A. Single Transmission Line Parameters Estimation Model

Based on the $\pi$-section model shown in Figure 1, one can find the relationship between the true PS voltages/currents and the line parameters as shown in (6).

$$\begin{bmatrix} V_i \\ V_j \end{bmatrix} = \mathbf{Z} \begin{bmatrix} I_i \\ I_j \end{bmatrix} \tag{6}$$

where, the impedance matrix is

$$\mathbf{Z} = \frac{1}{y(2+Zy)} \begin{bmatrix} 1+Zy & 1 \\ 1 & 1+Zy \end{bmatrix} \tag{7}$$

For the simplicity of the expression, the correction factors $KV_i$, $KI_i$, $KV_j$, and $KI_j$ are introduced. These variables are the reciprocals of the respective ratio errors and thus are equivalent. Using the products of the CT/PT measurements and correction factors to substitute the true values, (6) can be reformulated to be (8).

$$\begin{bmatrix} KV_i \times V_{iM} \\ KV_j \times V_{jM} \end{bmatrix} = \mathbf{Z} \cdot \begin{bmatrix} KI_i \times I_{iM} \\ KI_j \times I_{jM} \end{bmatrix} \tag{8}$$

Here, the 'ideal' PS measurements $V_{iM}$, $I_{iM}$, $V_{jM}$, and $I_{jM}$ are the values acquired by the transformation of direct three-phase measurement. As mentioned before, the ADCs in PMUs are designed to quantize the three-phase measurements before conducting this transformation. Therefore, the actual PS data provided by PMUs are transformed from quantized three-phase measurements. The differences or namely the PS quantization errors are

$$\begin{cases} \varepsilon_{V_{iM}} = V_{iM} - V_{iM,q} \\ \varepsilon_{I_{iM}} = I_{iM} - I_{iM,q} \end{cases} \tag{9}$$

Therefore, by introducing a new assisting variable $W = 1 + Z \cdot y$, (8) can be written as

$$\begin{bmatrix} V_{iM,q} + \varepsilon_{V_{iM}} \\ V_{jM,q} + \varepsilon_{V_{jM}} \end{bmatrix} = \hat{\mathbf{Z}} \cdot \begin{bmatrix} I_{iM,q} + \varepsilon_{I_{iM}} \\ I_{jM,q} + \varepsilon_{I_{jM}} \end{bmatrix} \tag{10}$$

where,

$$\hat{\mathbf{Z}} = \frac{1}{y(W+1)} \begin{bmatrix} W \dfrac{KI_i}{KV_i} & \dfrac{KI_j}{KV_i} \\ \dfrac{KI_i}{KV_j} & W \dfrac{KI_j}{KV_j} \end{bmatrix} \tag{11}$$

By reformulating (10), one has

$$\begin{bmatrix} V_{iM,q} \\ V_{jM,q} \end{bmatrix} = \hat{\mathbf{Z}} \cdot \begin{bmatrix} I_{iM,q} \\ I_{jM,q} \end{bmatrix} \underbrace{- \begin{bmatrix} \varepsilon_{V_{iM}} \\ \varepsilon_{V_{jM}} \end{bmatrix} + \hat{\mathbf{Z}} \cdot \begin{bmatrix} \varepsilon_{I_{iM}} \\ \varepsilon_{I_{jM}} \end{bmatrix}}_{\varepsilon} \tag{12}$$

Based on the previous analysis, the quantization errors are unbiased. Therefore, the expectation of the quantization errors part, $\varepsilon$, in the RHS of (12) should be a zero vector, which means

$$mean(\varepsilon) \to -\begin{bmatrix} E[\varepsilon_{V_{iM}}] \\ E[\varepsilon_{V_{jM}}] \end{bmatrix} + \hat{\mathbf{Z}} \cdot \begin{bmatrix} E[\varepsilon_{I_{iM}}] \\ E[\varepsilon_{I_{jM}}] \end{bmatrix} = \begin{bmatrix} 0 \\ 0 \end{bmatrix} \tag{13}$$

Thus, with adequate measurements to eliminate the quantization errors, the linear estimation problem can be formed as shown in (14).

$$\begin{bmatrix} \mathbf{I}_{iM,q} & \mathbf{I}_{jM,q} \end{bmatrix} \cdot \hat{\mathbf{Z}}^{\mathrm{T}} = \begin{bmatrix} \mathbf{V}_{iM,q} & \mathbf{V}_{jM,q} \end{bmatrix} - \varepsilon^{\mathrm{T}} \tag{14}$$

where, $\mathbf{V}_{iM,q} = \begin{bmatrix} V_{iM,q}^{(1)} & V_{iM,q}^{(2)} & \cdots & V_{iM,q}^{(N)} \end{bmatrix}^{\mathrm{T}}$, which is the vector of $N$ time-stamped, quantized voltage measurements provided by the PMU installed on bus $i$. The $\mathbf{V}_{jM,q}$, $\mathbf{I}_{iM,q}$, and $\mathbf{I}_{jM,q}$ are similar.

Using LSE and the single line parameters model formed in (14), one can get the estimated impedance matrix:

$$\hat{\mathbf{Z}}^{\mathrm{T}} = \left(\mathbf{I}_{M,q}^{\mathrm{T}} \mathbf{I}_{M,q}\right)^{-1} \mathbf{I}_{M,q}^{\mathrm{T}} \mathbf{V}_{M,q} - \left(\mathbf{I}_{M,q}^{\mathrm{T}} \mathbf{I}_{M,q}\right)^{-1} \mathbf{I}_{M,q}^{\mathrm{T}} \varepsilon^{\mathrm{T}} \tag{15}$$

where, $\mathbf{I}_{M,q} = \begin{bmatrix} \mathbf{I}_{iM,q} & \mathbf{I}_{jM,q} \end{bmatrix}$ and $\mathbf{V}_{M,q} = \begin{bmatrix} \mathbf{V}_{iM,q} & \mathbf{V}_{jM,q} \end{bmatrix}$.

Therefore, multiple times of LSE are to be performed using different sets of the original PMU data, instead of using the complete dataset at once. Taking advantage of the error's unbiased characteristic, the average of the multiple estimation results is used to eliminate the second part in RHS of (15). The final estimation results can be reached using only the quantized data, as shown in (16).

$$\hat{\mathbf{Z}}_{final}^{\mathrm{T}} = \sum_{k=1}^{P} \hat{\mathbf{Z}}_k^{\mathrm{T}} / P \tag{16}$$

where,

$$\hat{\mathbf{Z}}_k^{\mathrm{T}} = \left[ \left(\mathbf{I}_{M,q}^{(i)}\right)^{\mathrm{T}} \mathbf{I}_{M,q}^{(i)} \right]^{-1} \left(\mathbf{I}_{M,q}^{(i)}\right)^{\mathrm{T}} \mathbf{V}_{M,q}^{(i)} \tag{17}$$

which is the estimation results based on the $k$-th portion of the data when the original data set is partitioned into $P$ portions.

With the entries of $\hat{\mathbf{Z}}_{final}$: $\hat{Z}_{11}$, $\hat{Z}_{12}$, $\hat{Z}_{21}$, and $\hat{Z}_{22}$, the estimated impedance and susceptance can be calculated by first computing the estimated $\hat{W}$:

$$\hat{W} = \sqrt{\left(\hat{Z}_{11} \times \hat{Z}_{22}\right) / \left(\hat{Z}_{21} \times \hat{Z}_{12}\right)} \tag{18}$$

Given the correction factors of the voltage and current measurements of the bus $i$, namely $KV_i$ and $KI_i$, the correction factors on bus $j$ can be estimated as follow.

$$\begin{cases} KV_j = \left(\dfrac{1}{\hat{W}}\right)\left(\dfrac{\hat{Z}_{11}}{\hat{Z}_{21}}\right) KV_i \\ KI_j = \hat{W}\left(\dfrac{\hat{Z}_{21}}{\hat{Z}_{12}}\right) KI_i \end{cases} \tag{19}$$

The transducers installed on bus $j$ are thus calibrated.

Then the estimated line impedance and susceptance can be calculated as shown in (20) and (21).

$$\hat{y} = \sqrt{\frac{1}{\det\left(\hat{Z}_{final}\right)} \cdot \frac{KI_i KI_j}{KV_i KV_j} \cdot \frac{\left(\hat{W}-1\right)}{\left(\hat{W}+1\right)}} \tag{20}$$

$$\hat{Z} = \frac{1}{\hat{y}}\left(\hat{W}-1\right) \tag{21}$$

The single line parameters are estimated. Notably, line susceptances should always be positive. Hence, in (20), the positive imaginary part of the square roots is taken as the estimated susceptance.

As found in the numerical experiments, one factor that affects the estimation accuracy is the lack of variance on the PMU measurements when collected within a continuous time period. Such numerical characteristics could easily result in singularity during the computation process. Here we used a technique to avoid this problem taking advantage of PMUs' high reporting rate. For instance, we use one-hour data to conduct the estimation. Only the data measured in the first second of each minute are used and combined to form the estimation dataset. The PMU output frame rate is set to be 30 frames/sec. So the dataset contains 1800 time instances in total. As for the data partitioning for the results averaging process, all the 1st measurement frames of each second are used to form the 1st portion, all the 2nd frames to form the 2nd portion, etc. Hence, there are 30 datasets in total for each TL and the estimation is conducted 30 times with 60 time frames each time. The time period settings for raw data collection used here is user-defined. This setting can be shorter for more prompt estimation results, as long as adequate measurements can be provided. Therefore, the data selection manner is only provided for reference and should not affect the practicality of the implementation in reality.

### B. Propagation Methodology
For one single TL, the correction factors of CTs and PTs on the "To bus" can be estimated using the aforementioned method. Since the current TL shares this "To bus" with its consecutive TLs, a method is developed to propagate the estimated correction factors to these TLs. Therefore, starting from the reference bus where a set of revenue CTs and PTs deployed, one can conduct the estimation on each TL sequentially throughout the interconnected system topology.

#### 1) VOLTAGE ACCURACY PROPAGATION
The voltages on one bus measured on different TLs connect to it should be the same. So if there are $n$ lines that are connected to bus $i$ as shown in Figure 4, the voltages measured by different PTs on those lines should satisfy (22).

$$V_i^{(true)} = KV_i^{(1)} \times V_{iM}^{(1)} = KV_i^{(2)} \times V_{iM}^{(2)} = \cdots = KV_i^{(n)} \times V_{iM}^{(n)} \tag{22}$$

where, the superscript '*true*' indicates the true value of the measurement; the other superscripts indicate which line the measurement or correction factor is corresponded to.

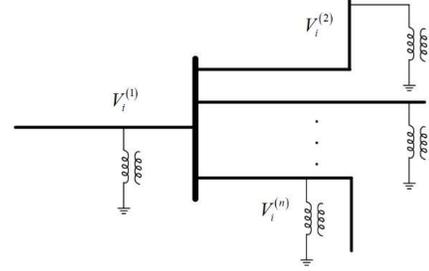

**Figure 4.** Voltage measurements on one single bus.

Considering the quantization effect, the quantized measurements can be found following this relationship:

$$V_i^{(true)} = KV_i^{(1)} \times \left(V_{iM,q}^{(1)} - \varepsilon_{V_{iM}}^{(1)}\right) = KV_i^{(2)} \times \left(V_{iM,q}^{(2)} - \varepsilon_{V_{iM}}^{(2)}\right)$$
$$= \cdots = KV_i^{(n)} \times \left(V_{iM,q}^{(n)} - \varepsilon_{V_{iM}}^{(n)}\right) \tag{23}$$

If one of the correction factors, $KV_i^{(j)}$, is known from the calibration of the previous line, for any of the other correction factors, $KV_i^{(k)}$, one has

$$KV_i^{(j)} \times V_{iM,q}^{(j)} = KV_i^{(k)} \times V_{iM,q}^{(k)} \underbrace{-KV_i^{(k)} \times \varepsilon_{V_{iM}}^{(k)} + KV_i^{(j)} \times \varepsilon_{V_{iM}}^{(j)}}_{\varepsilon_j} \tag{24}$$

Because of the unbiased characteristic of the quantization error, i.e. $\mathrm{E}\left[\varepsilon_{j'}\right] = 0$, one can use a similar technique as shown in subsection A to find $KV_i^{(k)}$.

#### 2) CURRENT ACCURACY PROPAGATION
Apart from the accurate voltage measurement, the accurate current measurements on the same bus also need to be propagated to form the reference bus for the consecutive line.

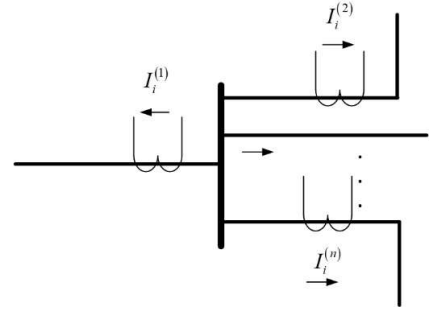

**Figure 5.** Injection currents measurements on one single bus.

Based on Kirchhoff's current law (KCL), the currents injections on the bus $i$ can be expressed as (25):

$$KI_i^{(1)} \times I_{iM}^{(1)} + KI_i^{(2)} \times I_{iM}^{(2)} + \cdots + KI_i^{(n)} \times I_{iM}^{(n)} = 0 \tag{25}$$

Given that one of the correction factors, $KI_i^{(1)}$, is known, (25) can be rewritten as (26). Such model forms another LSE

problem.

$$\begin{bmatrix} I_{iM}^{(2)} & I_{iM}^{(3)} & \cdots & I_{iM}^{(n)} \end{bmatrix} \begin{bmatrix} KI_i^{(2)} \\ KI_i^{(3)} \\ \vdots \\ KI_i^{(n)} \end{bmatrix}_{(n-1)\times 1} = \begin{bmatrix} -KI_i^{(1)} \cdot I_{iM}^{(1)} \end{bmatrix} \quad (26)$$

Because of the linearity of the estimation problem and the unbiased characteristic of the quantization errors, an averaging method similar to the aforementioned can be used to eliminate these errors. Therefore, one can directly reformulate (26) to be (27) and use different portions of the original quantized data to conduct the following optimization process. The final results will be the average of the optimums.

$$\begin{bmatrix} I_{iM,q}^{(2)} & I_{iM,q}^{(3)} & \cdots & I_{iM,q}^{(n)} \end{bmatrix} \begin{bmatrix} KI_i^{(2)} \\ KI_i^{(3)} \\ \vdots \\ KI_i^{(n)} \end{bmatrix}_{(n-1)\times 1} = \begin{bmatrix} -KI_i^{(1)} \cdot I_{iM,q}^{(1)} \end{bmatrix} \quad (27)$$

As stated in Section II. B. 1), the ratio errors have limits on magnitude and angle. Therefore, the correction factors' values should be bounded constrained. These constraints are in the form of magnitude and angles. But within the computation process, the ratio errors are calculated in the form of complex numbers. The constraints are projected from the polar coordinate plane into the complex plane and transferred from three single-phases into PS. The calculated constraints of the correction factors are shown in (28).

$$\begin{cases} K_{real} \in [0.9452, \ 1.0526] \\ K_{imag} \in [-0.1005, \ 0.1005] \end{cases} \quad (28)$$

However, the constraints derived are in the real plane, while the estimation model as shown in (26) is in the complex plane. In order to take these limits into consideration, the model is reformed as (29), (30), and (31).

$$\mathbf{I}_{reshape} = \begin{bmatrix} real\left(I_{iM,q}^{(2)}\right), imag\left(I_{iM,q}^{(2)}\right), \cdots, real\left(I_{iM,q}^{(n)}\right), imag\left(I_{iM,q}^{(n)}\right) \end{bmatrix} \quad (29)$$

$$\mathbf{I}_{reshape} \begin{bmatrix} real\left(KI_i^{(2)}\right), -imag\left(KI_i^{(2)}\right), \cdots, real\left(KI_i^{(n)}\right), -imag\left(KI_i^{(n)}\right) \end{bmatrix}^{\mathrm{T}}$$
$$= \begin{bmatrix} real\left(-KI_i^{(1)} \cdot I_{iM,q}^{(1)}\right) \end{bmatrix} \quad (30)$$

$$\mathbf{I}_{reshape} \begin{bmatrix} imag\left(KI_i^{(2)}\right), real\left(KI_i^{(2)}\right), \cdots, imag\left(KI_i^{(n)}\right), real\left(KI_i^{(n)}\right) \end{bmatrix}^{\mathrm{T}}$$
$$= \begin{bmatrix} imag\left(-KI_i^{(1)} \cdot I_{iM,q}^{(1)}\right) \end{bmatrix} \quad (31)$$

where, $real(\cdot)$ and $imag(\cdot)$ indicate the real and imaginary parts of the concerned variables; $KI_i^{(1)} \cdot I_{iM,q}^{(1)}$ is the calibrated current measurements.

The propagation process can be conducted on the whole transmission system with a mesh network structure. Chances are that multiple lines connecting the current bus being estimated and the currents flowing through them out of the bus are known. Under these circumstances, the known currents can be simply moved to the right-hand side of the equations. So that all the unknown correction factors of the other lines can be estimated.

Cases are also that there are double parallel lines or even triple parallel lines connecting two buses. There is a large probability that the currents flowing through such multiple lines are similar or even the same due to the same material and length of the line. The differences among measurements are small or close to zero, which severely influences the estimation accuracy. For example, if $I_{iM,q}^{(2)}$ and $I_{iM,q}^{(3)}$ are the two currents flowing out of bus $i$ through line 2 and line 3 respectively and those two lines are two parallel lines connecting bus $i$ and bus $j$, then the true values of those two currents are the same. Without pre-processing, even with constraints on the correction factors, the least squares optimization results could possibly be that $KI_i^{(2)}$ is around $1.0526 + j0.1005$ and $KI_i^{(3)}$ is around $0.9452 - j0.1005$. The correction factors' values are pushed to the limits and the accurate results are covered. Furthermore, part of the contribution of $I_{iM,q}^{(3)}$ to $I_{iM,q}^{(1)}$ is counted as $I_{iM,q}^{(3)}$'s. The solution to this issue is finding all the measurements of parallel lines and sum them together. The correction factors will also be summed up to one. The estimation process, in this case, is not aimed to find each accurate current but to find one accurate summation. Theoretically, with the same length, material, and operation environment, the currents flowing through those parallel lines should be the same. Therefore, with the accurate summation divide by the number of parallel lines, the accurate current flowing through one line can be reached. With the measurements on other lines, the correction factors can be calculated by taking the ratio of the true value and measurements respectively.

The estimation model formed previously can be simply transferred into a typical quadratic optimization problem with the bounded constraints listed in (28). The problem can be expressed as follows:

$$\min \ \left\| \mathbf{I}_{reshape} \mathbf{A}_1 \mathbf{k} - \mathbf{I}_{real}^{(true)} \right\|_2^2 + \left\| \mathbf{I}_{reshape} \mathbf{A}_2 \mathbf{k} - \mathbf{I}_{imag}^{(true)} \right\|_2^2$$
$$s.t. \quad l_{real} \le \mathbf{k}_{real} \le u_{real} \quad (32)$$
$$l_{imag} \le \mathbf{k}_{imag} \le u_{imag}$$

where,

$$\mathbf{k} = \begin{bmatrix} real\left(KI_i^{(2)}\right), imag\left(KI_i^{(2)}\right), \cdots, real\left(KI_i^{(n)}\right), imag\left(KI_i^{(n)}\right) \end{bmatrix} \quad (33)$$

$$\mathbf{A}_1 = \begin{bmatrix} \mathbf{B}_1 & & \mathbf{0} \\ & \ddots & \\ \mathbf{0} & & \mathbf{B}_1 \end{bmatrix}, \quad \mathbf{B}_1 = \begin{bmatrix} 1 & 0 \\ 0 & -1 \end{bmatrix}, \quad \mathbf{A}_2 = \begin{bmatrix} \mathbf{B}_2 & & \mathbf{0} \\ & \ddots & \\ \mathbf{0} & & \mathbf{B}_2 \end{bmatrix}, \quad \mathbf{B}_2 = \begin{bmatrix} 0 & 1 \\ 1 & 0 \end{bmatrix} \quad (34)$$

$\mathbf{k}_{real} = \begin{bmatrix} \cdots & K_{real}^{(j)} & \cdots \end{bmatrix}$ and $\mathbf{k}_{imag} = \begin{bmatrix} \cdots & K_{imag}^{(j)} & \cdots \end{bmatrix}$ are vectors that contain real and imaginary parts of the correction factors, respectively; $\mathbf{A}_1$, $\mathbf{B}_1$, $\mathbf{A}_2$, and $\mathbf{B}_2$ are constant parameter matrices assisting to vectorise the problem.

With a typical bounded constrained quadratic optimization method, the correction factors of all the other lines connected to the concerned bus can be estimated. Provided the knowledge of these propagated correction factors, the reference of the consecutive TL is found and the estimation can be performed.

## C. System Topology

Based on the former propagation methods, the accuracy of the reference bus can be passed to the consecutive lines through the interconnected topology of the system.

Given that transmission systems usually form mesh-structured networks, the Edge-based Breadth-first Search (EBBFS) algorithm is utilized in this work to construct the tree topology with the reference bus as the root. The flow chart of this algorithm can be found in Figure 6. Unlike the traditional Breadth-first Search algorithm trying to find node visiting sequence, the proposed method is more focused on visiting the edges/lines in the network. The nodes/buses, on the other hand, can be visited multiple times due to the mesh structure to make sure that all the lines in the network are visited. Therefore, this EBBFS method is more suitable to the line parameter estimation problem under discussion.

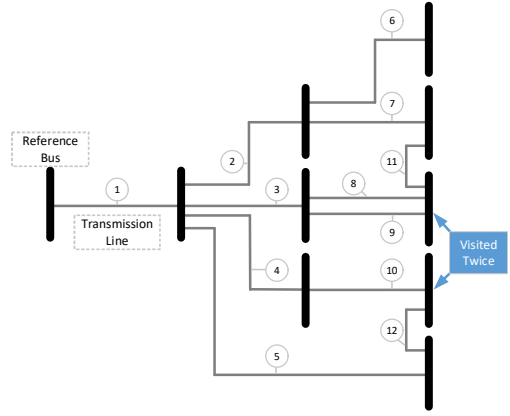

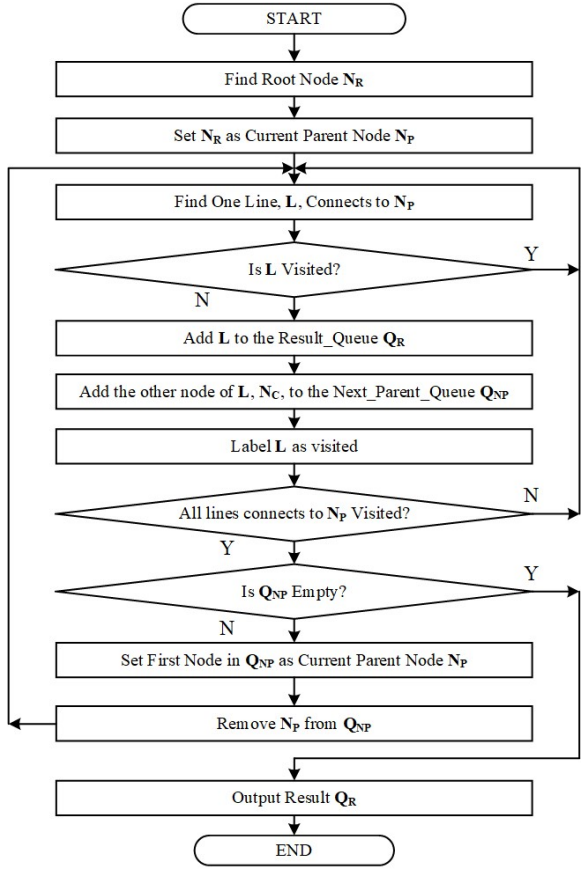

**Figure 6.** The edge-based breadth-first search algorithm flowchart.

**Figure 7.** The edge-based breadth-first search traverse order.

first search keeps the tree's height as small as possible. So that the accuracy has the slowest decay rate.

The flow chart in Figure 8 demonstrates the complete system PS parameters estimation and instrument transformers calibration process.

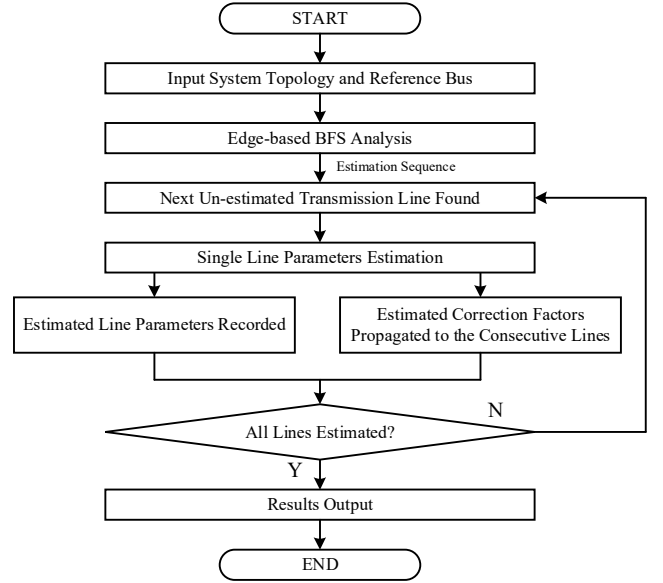

**Figure 8.** The complete process of the line parameter estimation.

The typical search process can be seen in Figure 7. The numbers in the small circles indicate the order in which the line is evaluated. As can be seen, there are two nodes/buses in the network were visited twice to make sure the 11th and 12th lines are estimated.

The EBBFS can traverse all the concerned TLs in the system. When one unvisited line is found by the algorithm, the correction factors of the *"From bus"* of this line are guaranteed to have been estimated during the processing of the previous lines. These methods can assure that all the concerned lines are analyzed. At the same time, the breadth-

## IV. CASE STUDY

Two test cases, the IEEE 118-bus system and Texas 2000-bus system [26], have been used to validate the proposed methodology. Synthetic data based on these cases are generated and used as the input of the estimation procedure. The reporting frame rate of simulated PMU is set to be 30 frames/second. Hence, the total number of time instances for one hour is $30 \times 60 \times 60 = 108,000$. Based on the strategy introduced in Section III. A, 1800 time instances of power flow are conducted based on the load curve of one summer day morning-load-pick-up hour data of the Dominion Energy Virginia system. The voltages and currents are recorded as the original true values.

PMU measurements are generalized based on the aforementioned error model. The three-phase ratio errors are randomly generated within the boundary of $r_{M,ph} \in [0.95, 1.05]$ and $r_{A,ph} \in [-5°, 5°]$. They will remain constant during the whole estimation process because of the limited time duration to be simulated. Although, according to the accuracy classes listed in the IEEE standard [5], the errors are far smaller in reality. We use larger values to demonstrate the effectiveness and robustness of the method here. The PS ratio errors are thus calculated with (4) and recorded as part of the true values to be estimated. The simulated three-phase measurements are created using the accurate three-phase voltages and currents obtained as well as the generated three-phase ratio errors. The simulated PS measurements are reached by quantizing these processed three-phase data. The quantization scale for voltages is 12 V and that for currents is 0.65 A.

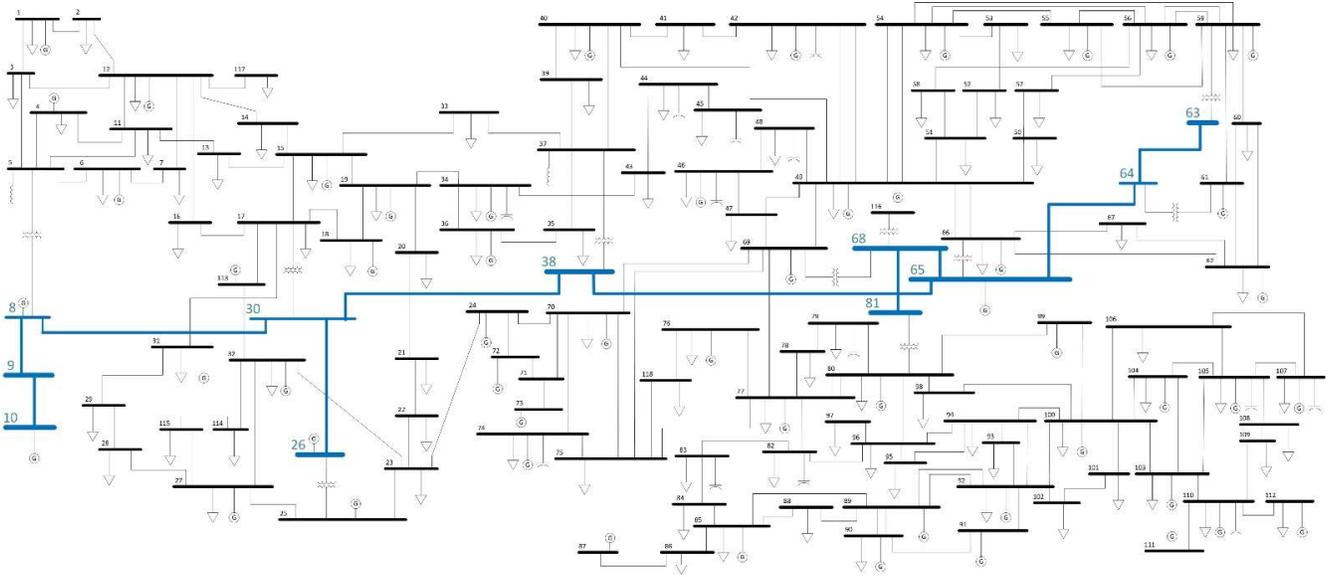

**Figure 9.** The IEEE 118-bus system diagram.

### A. IEEE 118-Bus System

#### 1) SYSTEM AND PMU MEASUREMENTS GENERATION

The IEEE 118-bus system diagram can be found in Figure 9. There are two voltage levels in the system: Buses 8, 9, 10, 26, 30, 38, 63, 64, 65, 68, 81 are 345 KV buses (blue and bold buses and lines in the system diagram); The others are 138 KV buses (black buses and lines). The analysis was conducted on the high-voltage-level (345 KV) buses. Notably, the methodology is also suitable to power systems of other voltage levels.

Based on this system, PS voltages and currents are acquired for all 345kV buses through AC power flow calculations. Even though estimation was only performed on the lines connecting the 345kV buses, the currents injections from lower voltage levels are also collected due to the data requirements of the current accuracy propagation method. All 24 relevant lines can be easily identified through the system diagram.

Without losing generality, Bus 81 is chosen as the reference bus and the line connecting Bus 81 and Bus 68 as the first TL to be estimated. This means the synthetic voltage and current measurements of Bus 81 on this line are the accurate ones. For other voltage and current measurements, the PMU measurement error model derived in Section II. B is deployed to simulate the measurements provided by the local PMUs with ratio errors.

For transposed lines, it is assumed that the system remains three-phase balanced over the operation time. The single-phase power flow results are taken as the Phase A data, and data in the other two phases are generated through ±120° phase angle shifts. One single un-transposed line is taken as an example to demonstrate the effectiveness of the estimation algorithm on this circumstance. The three-phase voltages and currents of the reference bus are generated similarly as the transposed case based on the power flow results. However, those of the other bus are calculated using the provided impedance and susceptance matrices [27].

#### 2) SINGLE LINE PARAMETERS ESTIMATION RESULTS

For transposed lines, the method developed in Section III. A is used to estimate the parameters of the line connecting Bus 81 and Bus 68. With accurate voltage and current measurements known on Bus 81, its correction factors are $KV_1 = KI_1 = 1 + j0$. The correction factors of the voltage on and the current flowing out of Bus 68, $KV_2$ and $KI_2$ are also estimated for instrument transformers calibration. The computation results are shown in TABLE I.

The estimation results indicate good accuracy. The respective error rates of the correction factors are 7.8956E-06 % for $KV_2$ real part, 7.6117E-02% for $KV_2$ imaginary part, 6.1495E-03% for $KI_2$ real part, and 1.5740% for $KI_2$ imaginary part. The errors of imaginary parts are larger than that of the corresponding real parts. The main reason is that the absolute values of the imaginary parts of the correction factors are much smaller than the real parts by nature. This is brought by the ratio errors angle limits defined by the standards. This numeric characteristic highly influenced the precision of estimating the imaginary parts. However, based on the same numeric feature, the accuracy of the line parameters estimation is remarkable, which is not interfered by the imaginary parts of the correction factors. The error rate of the resistance, reactance, and susceptance estimation are 6.1517E-02%, 7.1649E-04%, and 6.5191E-02% respectively.

| | | | True Value (× 10⁻⁷) | Estimated Value (× 10⁻⁷) | Error (× 10⁻⁷) |
|---|---|---|---|---|---|
| Bus 1 | 81 | $KV_1$ | $1 \times 10^7 + 0i$ | N/A | N/A |
| | | $KI_1$ | $1 \times 10^7 + 0i$ | N/A | N/A |
| Bus 2 | 68 | $KV_2$ | 10266464.9898 -19846.0026i | 10266465.8004 -19861.1088i | -0.8106 -15.1062i |
| | | $KI_2$ | 9932573.4041 +115061.5151i | 9931962.5986 +116872.5726i | 610.8055 +1811.0575i |
| R (pu) | | | 17500.0044 | 17510.7698 | -10.7654 |
| X (pu) | | | 202000.0090 | 202001.4563 | -1.4473 |
| y (pu) | | | 4040000.1451 | 4037366.4274 | -2633.7177 |

**TABLE I**
**SINGLE TRANSPOSED LINE PARAMETERS ESTIMATION RESULTS**

For the un-transposed line, the impedance matrix and the susceptance matrix used to generate three-phase data are as follows:

$$\mathbf{Z}_{abc} = \begin{bmatrix} 8.5922 + 61.0128i & 4.1208 + 27.6955i & 4.0940 + 23.9696i \\ 4.1208 + 27.6955i & 8.5131 + 61.0865i & 4.0932 + 27.7693i \\ 4.0940 + 23.9696i & 4.0932 + 27.7693i & 8.4612 + 61.1171i \end{bmatrix} \tag{35}$$

$$\mathbf{y}_{abc} = \begin{bmatrix} 1.0913\text{E-}04i & -1.6076\text{E-}05i & -1.6076\text{E-}05i \\ -1.6076\text{E-}05i & 1.0544\text{E-}04i & -2.4244\text{E-}05i \\ -1.6076\text{E-}05i & -2.4244\text{E-}05i & 1.0544\text{E-}04i \end{bmatrix} \tag{36}$$

The estimation results are listed in TABLE II. The absolute error rates of the PS three line parameters are all smaller than 2%. For the most important parameter, the PS line reactance, the error rate is smaller than 0.5%. Even though the overall accuracy of the estimation is lower than the transposed case, it is still satisfactory. These results clearly demonstrate the effectiveness of the single TL parameters estimation method when deployed to both transposed and un-transposed TLs.

**TABLE II**
**SINGLE UN-TRANSPOSED LINE PARAMETERS ESTIMATION RESULTS**

| Positive Sequence Parameters | True Value | Estimated Value | Error | Error Rate (%) |
|---|---|---|---|---|
| $R(\Omega)$ | 4.4195 | 4.3377 | -0.0818 | -1.8509 |
| $X(\Omega)$ | 34.5940 | 34.7055 | 0.1115 | 0.3223 |
| $y(S)$ | 1.2547E-04 | 1.2375E-04 | -1.7155E-06 | -1.3708 |

### 3) SYSTEM LINES PARAMETERS ESTIMATION

Since the estimation accuracy difference between the transposed and un-transposed lines is small, the balanced system cases are utilized to illustrate the capability of the proposed algorithm for simplicity. Given the effectiveness of the single line parameter estimation algorithm, with the methods in Section III. B, the estimation process is propagated towards the 345 KV subsystem. The subsystem diagram is shown in Figure 10. The numbers in circles indicate bus numbers. Propagation order provided by EBBFS is from left to right. The reference bus and line remain the same as in the single line estimation process case.

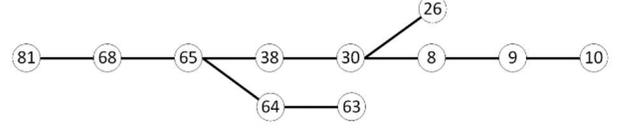

**Figure 10.** The 345 KV subsystem diagram.

The estimation results are shown in TABLE III. It can be seen that the estimation is generally accurate. Most of the error rates of the estimated line parameters are smaller than or around 1%. However, comparatively notable errors are also present. The error rate of the resistance and susceptance of the line between Bus 64 and Bus 63 are 12.8980% and 8.6431% respectively.

The possible reasons for this phenomenon include: 1) this line is a shorter TL compared with others. The susceptance of this line is far smaller than the other lines, which can cause numerical errors within (20). Because the line impedance is estimated based on the knowledge of susceptance as shown in (21), the error of susceptance estimation can be easily passed to the line impedance. 2) this line is comparatively far from the reference bus. Such distance can result in the accumulative effect of the ratio errors, due to the propagation processes. Without the propagation of accurate measurements, the estimation process of this line can be easily influenced. One can also observe around 2% error rates of the estimation of the lines connecting Bus 30 with Bus 26, Bus 8 with Bus 9, and Bus 9 with Bus 10. These larger error rates are likewise brought by the distances from the reference bus. The table is arranged based on the visited order of the lines in the subsystem. A pattern can be summarized indicating the increase of the errors along with the consecutive visits of the lines.

### B. Texas 2000-Bus System

The Texas 2000-bus system, available in [26] contains 2007 buses and 2481 lines with three voltage levels: 13.8KV, 115KV, and 345KV. The 345KV subsystem is composed of 224 buses and interconnected by 335 lines. The voltage and current data were generalized in the same manner as described in subsection A. 1). Only one bus is selected as the reference bus and equipped with accurate voltage and current data.

The estimation results using the three-phase balanced data are shown in the following figures. For the reactances estimation results are shown in Figure 11, most of the lines' error rates are within [-4%, 2%] and the largest does not exceed -7%. For the susceptances, most of the lines' error rates lay in the range of [-5%, 5%] as shown in Figure 12. This shows that even for the large scale systems with more

| From Bus | KV$_1$ Error | KI$_1$ Error | To Bus | KV$_2$ Error | KI$_2$ Error | R Error | R Error Rate (%) | X Error | X Error Rate (%) | y Error | y Error Rate (%) |
|---|---|---|---|---|---|---|---|---|---|---|---|
| 81 | N/A | N/A | 68 | -8.1052E-08 -1.5106E-06i | 6.1081E-05 -1.8111E-04i | 1.08E-06 | 0.0616 | 1.44E-07 | 0.0007 | 2.63E-04 | 0.0652 |
| 68 | 1.8552E-07 +9.9730E-07i | -2.5809E-03 2.2055E-02i | 65 | -1.8756E-06 +4.4313E-07i | -1.2458E-03 +2.1474E-02i | 3.55E-06 | 0.2570 | 2.59E-05 | 0.1619 | 0.00033 | 0.1027 |
| 65 | -2.3162E-06 +6.4019E-07i | 9.7546E-03 +3.9078E-02i | 38 | 4.3103E-05 -8.7327E-05i | 1.2062E-02 +4.1090E-02i | 1.39E-04 | 1.5530 | 9.95E-04 | 1.0088 | 0.00828 | 1.5832 |
| 65 | -1.8341E-06 +3.2975E-07i | -1.5036E-02 +2.2295E-02i | 64 | 7.3974E-06 -3.6754E-06i | -1.5872E-02 -5.5203E-02i | 1.16E-05 | 0.4321 | 4.61E-04 | 1.5270 | 0.00088 | 0.4645 |
| 38 | 4.6445E-05 -8.3559E-05i | 1.2282E-02 -4.0879E-02i | 30 | 4.0532E-05 -7.9476E-05i | 1.1226E-02 +4.1224E-02i | 5.59E-05 | 1.2041 | 5.51E-04 | 1.0204 | 0.00251 | 1.1922 |
| 64 | 7.0899E-06 -4.1999E-06i | -1.6471E-02 +2.4791E-02i | 63 | 6.8327E-06 -8.7721E-05i | 2.3150E-02 +4.0754E-02i | 2.22E-04 | 12.8980 | 4.19E-06 | 2.0957 | 0.00933 | 8.6431 |
| 30 | 4.2255E-05 -7.4048E-05i | 1.6471E-03 +5.7800E-02i | 8 | 7.6776E-05 +1.0365E-04i | 4.8530E-03 +5.7460E-02i | 6.33E-05 | 1.4697 | 1.54E-04 | 0.3052 | 0.00024 | 0.0948 |
| 30 | 3.8312E-05 -7.5784E-05i | 3.2837E-02 +1.5116E-02i | 26 | 1.5661E-04 +1.0000E-05i | 3.2773E-02 -1.3359E-02i | 3.71E-04 | 4.6407 | 2.86E-03 | 3.3271 | 0.01675 | 3.6905 |
| 8 | 7.4647E-05 +1.0822E-04i | 2.2547E-02 +7.9627E-03i | 9 | 5.0781E-05 -8.8060E-05i | 2.2142E-02 -7.4259E-03i | 7.21E-05 | 2.9545 | 7.08E-04 | 2.2314 | 0.01436 | 2.4729 |
| 19 | 5.1020E-05 -8.9386E-05i | 2.2917E-02 -6.3834E-03i | 10 | 4.4876E-05 -1.0376E-04i | 3.3270E-02 -6.0184E-03i | 4.88E-05 | 1.8929 | 7.59E-04 | 2.3561 | 0.01551 | 2.5224 |

than 300 TLs to be considered, the proposed method is still effective. Besides, there is much potential for estimation accuracy improvement with more references included. As a matter of fact, for large scale systems, there are more likely to exist multiple tie lines and most of these lines are equipped with revenue CTs and PTs that are suitable to be taken as references in this method.

However, there do exist several cases where the error rates are larger than 10% for susceptances. The comparison of susceptances values is made between the lines with larger error rates and the whole group of the lines. The box plots are shown in Figure 13. The left box plot shows the distribution of the susceptances of lines with more than 5% estimation error rates. The right one demonstrates the distribution of all the 335 susceptances within the concerned 345KV subsystem. As one can see in the figure, the lines with larger estimation errors generally are far smaller than the other lines. This confirms the reason described in section 3) that shorter TLs with smaller parameters might influence the estimation accuracy.

The possible solution to the comparatively large estimation errors can be obtained by integrating more reference measurements into the system. Especially for those short TLs that have significantly small parameters, installation of the revenue CTs and PTs will effectively relieve them of the cumulative errors propagated from former lines and provide more accurate initial data for estimation. Nonetheless, the installation will also increase

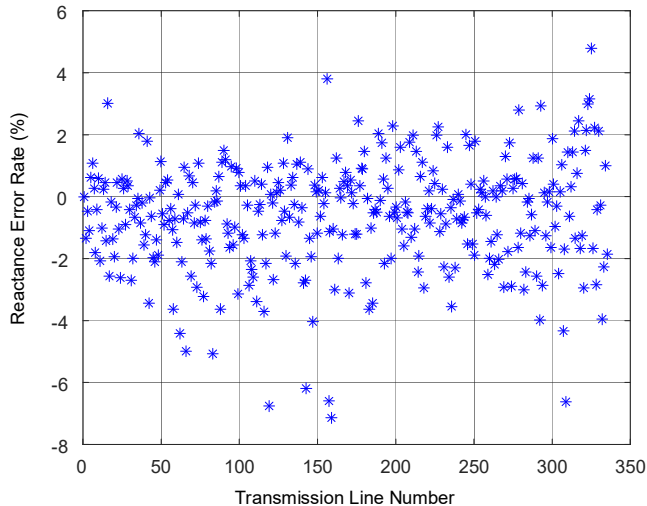

**Figure 11.** Estimation error rate of reactances in Texas 2000-bus system.

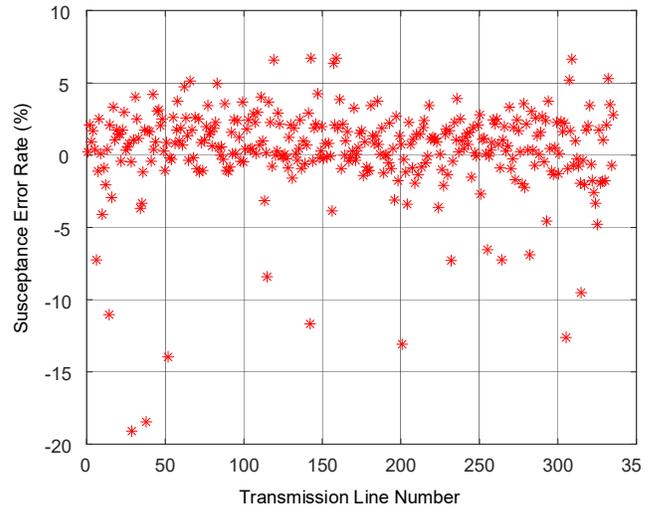

**Figure 12.** Estimation error rate of susceptances in Texas 2000-bus system.

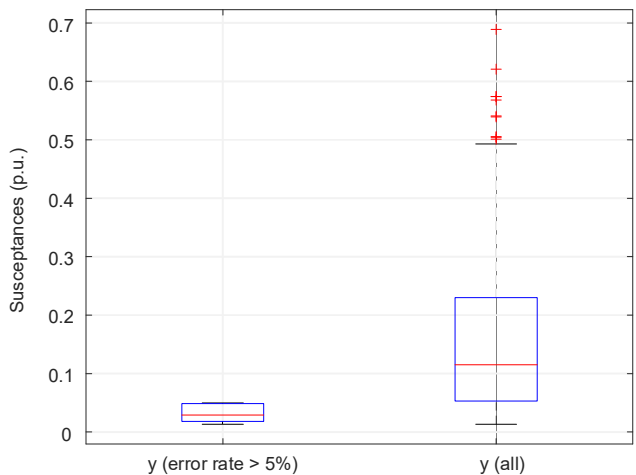

**Figure 13. Comparison of the distribution of lines susceptances in Texas 2000-bus system.**

the economic cost.

Although the estimation results of several parameters are comparatively large, most of the error rates are limited to be within -3% and 3%. This level of accuracy demonstrates that the system lines parameters estimation methodology is effective for most cases. At the same time, the adaptiveness of the algorithm is also demonstrated. With proper measurements and system topology information, the method is applicable to any other power systems.

## V. CONCLUSION

In this paper, a methodology is proposed to estimate transposed/un-transposed TL PS parameters and calibrate instrument transformers in a power system. The method is based on the proposed PMU measurement error model. With the reference bus selected, a novel estimation scheme is utilized to deal with the quantization characteristics of PMU measurements. The estimated accuracy is then propagated to the whole system following the network topology acquired by the Edge-based Breadth-first Search method. An advanced current measurements propagation method based on quadratic optimization is introduced in this propagation process to consider ratio errors boundary and to overcome the parallel lines' influence.

Both the single TL and the system lines parameters estimation methods are verified through the IEEE 118-bus standard system. The algorithm is also utilized to larger-scale Texas 2000-bus system. The results show satisfactory accuracy. Most of the estimation errors are around 1%. Very few of them are comparatively large to reach 3% to 4% for the 118-bus system and more than 10% for the 2000-bus system. The possible reasons are also proposed and analyzed, including short TLs with exceptional small parameters and accuracy decaying with large scale systems. Both reveal the potential for further investigation. However, the methodology proposed is effective and based entirely on the software which can save utilities much fieldwork efforts. With this method, utilities can acquire system status very promptly with low economic and time cost. The line parameters and transducers' correction factors estimated can be used for further grid applications in wider areas. Extension to other systems is also straightforward with sufficient PMU measurements and system topology information. Given the PS data acquisition reality during common utility operation, this method can satisfy most utilities' requirements and therefore can be widely deployed. As of the completion of this research article, relevant open-source application (available at [28]) has been developed based on the openECA platform (PMU data concentrator) provided by the Grid Protection Alliance.